\documentclass[twocolumn, showpacs, amsmath, amssymb, aps, prl]{revtex4-1}
\usepackage{graphicx}
\usepackage{dcolumn}
\usepackage{bm}

\begin{document}

\title{Quantum flights}

\author{Evgeny\,G.\,Fateev}

 \email{e.g.fateev@gmail.com}
\affiliation{%
Institute of mechanics, Ural Branch of the RAS, Izhevsk 426067, Russia
}%
\date{\today}

\begin{abstract}
The principles of quantum motors based on Casimir platforms (thin-film 
nanostructures are at issue) are discussed in plain language. The generation 
of quantum propulsion is caused by the noncompensated integral action of 
virtual photon momenta upon a configuration unit cell in the platform. The 
cells in a Casimir platform should be situated in a certain order with 
optimal geometric parameters. The evaluation of the quantum propulsion shows 
that, for example, ten square meters of ideal Casimir platforms (it is a 
complex single-layer structure) could make Cheops pyramid move!
\end{abstract}

                           
\maketitle

\textbf{INTRODUCTION}

Quite recently the idea of the creation of Casimir platforms has been 
suggested \cite{Fateev:2013}. They can be the nucleus of universal 
quantum propulsion devices in the nearest future. What is at issue is a far 
more serious and universal quantum effect than quantum levitation 
\cite{Leonhardt:2007} (the hovering of one body over another). The 
platform motion is based on the Casimir expulsion effect 
\cite{Fateev:2012} which can make it possible that people would fly 
wherever they want with whatever speeds even with the speed of light; and 
not only fly but perform a great variety of useful and wonderful actions 
(without any magic) \cite{Fateev:2013}. 
\begin{figure*}
\hypertarget{fig1}
\centerline{
\includegraphics[width=6.0in,height=2.5in,keepaspectratio]{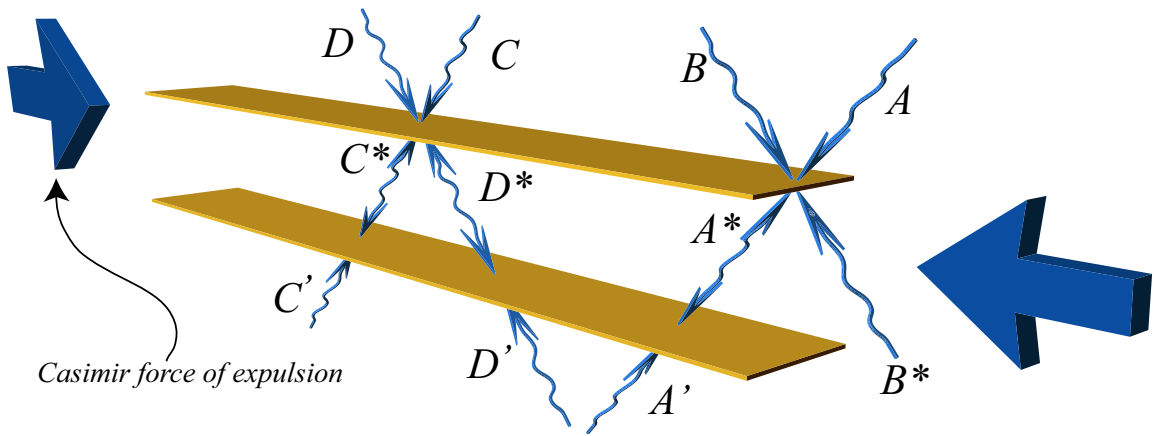}
\label{fig1}
\caption{The scheme with two very thin plates (wings); the arrows imitate the 
rays with virtual photons. The projections of the rays $D$, $C$ and $D^\ast 
$, $C^\ast $ upon the plates compensate one another. The projections of the 
rays $B$, $A$ and $B^\ast $, $A^\ast $ at the ends of the plates are not 
compensated \cite{Casimir:1948}. As a result, the Casimir expulsion 
force \cite{Fateev:2012} acts upon the plates from both ends tending 
to make ``washboards'' from the parallel plates.}}
\end{figure*}
Strictly speaking, the platforms can be referred to non-fuel motors. It does 
not mean that they can be called ``perpetuum mobiles''; however, there is 
something perpetual in them. What is the difference? According to the 
definition \cite{Stanley:1968} a perpetuum 
mobile is ``an imaginary device allowing obtaining useful work larger than 
the energy given to it''. The Casimir platforms get the energy from virtual 
photons continuously appearing in physical vacuum. 

However, will the 
platforms perform more work than they get the energy from the ``boiling'' 
physical vacuum? Of course, they will not. How will the energy, which we 
decide to use for the Casimir platforms, be transformed? The answer is 
trivial: it will be transformed into kinetic energy and heat.

Another question that can arise is: what about the law of momentum 
conservation in the system? Remember that the above law is valid only for 
closed systems. A system is called a closed system when no external forces 
act upon it or when the action of the above forces is compensated. We have a 
system which is open to all the virtual photon ``winds''. Therefore, photons 
will give their momenta to our wings for an indefinitely long period.

And since the energy of the Universe (of the large known to us part of the 
Universe) is rather large and is not consumed yet (by anyone or anything), 
using the Casimir platforms we ourselves can quickly bring the Universe to 
heat death. However, let us not worry before the time comes; the energy is 
free but we need to spend it economically like oil and gas. Will we be able 
to save it? For how long will it last? If such questions arise it means that 
we are discussing a limited source of energy with the capacity of several 
billion years and not a perpetuum mobile.

And why is the Universe the source of limited energy? Here we should agree 
with Steven Hocking \cite{Hawking:2010} that we just live in a certain 
fluctuation or, perhaps, ``layer'' of the space where a ``heap'' of positive 
energy has appeared and at some other place there is a ''hole'' of negative 
energy. Anyway, the sum of the energies is zero. And we are going to use our 
``heap'' of energy until it itself disappears in some ``quantum jump''. How 
can it be done? Recently I have understood that it is not very complicated. 
Here I will try to explain the principle of the performance of such 
platforms in details without using formulae as far as possible. Those who 
wish may read articles \cite{Fateev:2012,Fateev:2012a,Fateev:2013a} and 
appreciate their simplicity.

\textbf{WHAT IS CASIMIR EXPULSION?}

Well, what are the Casimir platforms and why should they perform in the way 
we want them to perform? Let me remind you here of something concerning 
classical Casimir effect. At present almost everyone knows that two 
perfectly conducting parallel metal plates should attract to one another if 
they are placed at nanometer distances relative to one another. This effect 
has already been found experimentally \cite{Lamoreaux:1997} and it proves 
that the mechanical action of virtual photons upon solid matter is real.

Developing the theory of the effect in 1948 \cite{Casimir:1948}, Hendrik 
Casimir made several approximations in the model. It was necessary at that 
time when nobody was dreaming of having a laptop with packages of 
mathematical programs for all cases of life. First, for the simplicity of 
the performance of his mathematical calculations Casimir made the 
approximation that the plates are strictly parallel and infinite. Second, he 
assumed that the incidence of virtual photons upon the plates is normal; and 
photons that are incident to the normal at angles mutually compensate their 
action in the direction parallel to the plates, because, for example, for 
any photon incident at any angle there is always a symmetrically incident 
photon. This suggestion refers both to external and internal virtual photons 
between the plates (see Fig.\hyperlink{fig1} 1). 
\begin{figure*}[htbp]
\hypertarget{fig2}
\centerline{
\includegraphics[width=2.2in,height=2.2in,keepaspectratio]{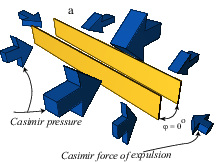}
\includegraphics[width=2.2in,height=2.2in,keepaspectratio]{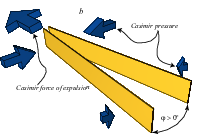}
\includegraphics[width=1.8in,height=1.5in,keepaspectratio]{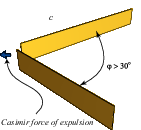}
\label{fig2}
\caption{The schemes with two very thin plates; the arrows imitate the Casimir 
pressure onto the surface (the value of the pressure on different areas of 
the plates is imitated by the sizes of the arrows) \cite{Fateev:2012}. 
The ends are also acted upon by the Casimir expulsion force. This force is 
completely compensated for the parallel configuration ($a$) 
\cite{Fateev:2012a}. In the nonparallel configuration ($b$) there is 
noncompensated expulsion force. At large angles, there is also 
non-compensation; however, it is very little ($c$).}}
\end{figure*}
Casimir also thought that at different frequencies the intensity of the 
photon production in vacuum was the same. In spite of the above 
approximations it is believed that Casimir's resultant formula is quite 
valid for the calculation of Casimir pressure between the parallel plates. 
There were several attempts to find Casimir pressure for the cases with 
nonparallel and finite plates; however they did not lead to any results 
which would be important and useful for further progress. The above attempts 
did not allow to find Casimir expulsion forces possible in such systems 
either.

The calculation of Casimir pressure values for more complex systems than 
parallel plates is still a very intricate problem. Even for joint 
configurations such as a ball and plane, two balls, two cones, a cone and 
plane, a trapezoid and plane, and others it is necessary to use rough 
approximations (see \cite{Milton:2001,Bordag:2009}); for example, 
the approximation by Deryagin \cite{Derjaguin:1956} which he developed in 
the 30s of the XXth century for Van der Waals forces between particles in 
liquid dispersed mixtures. The concept of the similarity of Casimir forces 
and Van der Waals forces is still used at present and it is even being 
improved for approximate calculations of Casimir forces for many 
configurations. Without mentioning the details of the Deryagin method, let 
us note that Van der Waals forces differ from Casimir forces significantly. 
Van der Waals forces act between inner opposite surface layers of figures. 
In contrast to them, Casimir forces are caused by the difference between the 
action of virtual photons upon external and inner surfaces of 
configurations.

The understanding of the drawbacks and roughness of the early calculation 
methods has allowed to develop some optical approximation for the 
calculation of Casimir forces. In this approximation, nonparallelism and 
finiteness of two plates (wings) and all possible angles of the incidence of 
virtual photons upon external and inner surfaces are taken into account. 
Scientists have succeeded in calculating the Casimir pressure along the 
entire length of plates and at the ends as well. Moreover, using the above 
approximation the Casimir expulsion force has been found, which is directed 
perpendicularly to the Casimir pressure (see Fig.\hyperlink{fig2} 2a).

The calculations result in finding amazing things. For example, it is found 
that Casimir forces tend to press parallel plates to one another and, 
moreover, to rumple them into ``washboards'' \cite{Fateev:2012}. It 
can happen due to the action of expulsion forces at the ends of the plates 
in the direction toward one another as in Fig.\hyperlink{fig2} 2a. Certainly it is not the 
most impressive manifestation of expulsion forces. In the case when the 
plates are slightly nonparallel, the expulsion forces at different ends will 
become unequal and opposite in their direction. As a result, the 
configuration from two plates fixed relative to one another will be 
influenced by noncompensated expulsion force as shown in Fig.\hyperlink{fig2} 2b. The 
magnitude of the force will depend on the angle $2\varphi $ between the plates. 
There are certain average angles, at which the expulsion force is maximal. 
In this case, for getting maximum of vacuum energy for the generation of 
expulsion forces, there is no need to make the plates infinitely long. There 
are optimal lengths $R$ for wings, and when the wings are made longer than 
those lengths, the expulsion forces in the configuration do not increase.

It turns out that everything found for a single configuration (see Fig.\hyperlink{fig3} 3a) 
is valid for periodically situated figures. A certain phenomenon 
\cite{Fateev:2012a} has been found; when plates are placed at 
absolutely equal distances $d=a$ from one another, there is no expulsion in 
the entire configuration (see Fig.\hyperlink{fig3} 3b). However, when the relation of the 
distances of the smallest sections of figures is different ($d/a>1$), at any 
angle $\varphi $, an expulsion force will appear in the entire configuration 
(Fig.\hyperlink{fig3} 3c). The force will be linearly dependent on the number of wings in a 
periodic configuration. It means that the force can be whatever we wish!

Let us also note that in a periodic system, for any angles $\varphi $ there is 
an optimal location parameter $d/a\sim 1.65$, at which the configuration 
expulsion force is maximal. Incidentally, this number almost coincides with 
the number 1.618 characterizing the ``golden section''. It means that there 
are three parameters, the use of which make it possible to optimize a 
periodic configuration, namely, the length of wings R, the angle between the 
wings $\varphi $, and the relation $d/a$.

\textbf{ESTIMATES OF EXPULSION FORCES}

Let us give simple estimates of geometric and physical parameters of the 
Casimir platforms, which are necessary for lifting, for example, Cheops 
pyramid. 
\begin{figure*}[htbp]
\hypertarget{fig3}
\centerline{
\includegraphics[width=2.0in,height=2.0in,keepaspectratio]{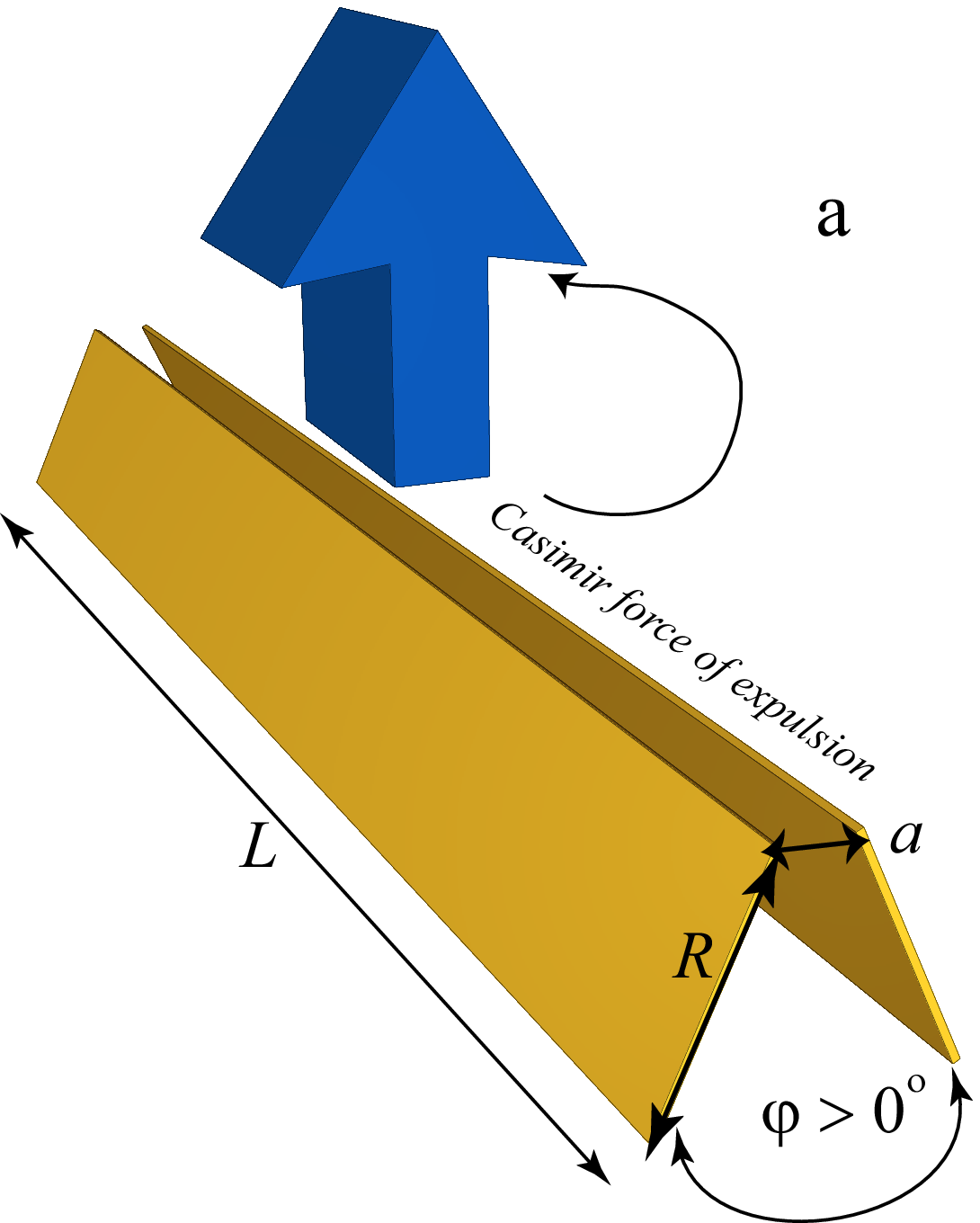}
\includegraphics[width=2.4in,height=2.4in,keepaspectratio]{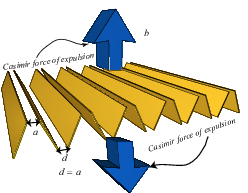}
\includegraphics[width=2.4in,height=2.4in,keepaspectratio]{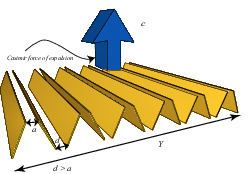}
\label{fig3}
\caption{The scheme with very thin plates with the length of the wings $R$, 
which are a little larger than the distance $a$, the width $L\sim 1$\,m and 
the angle $\varphi $ between them. For such configuration there is always 
expulsion force directed to the side of deceasing section of a cavity ($a)$. 
For the periodic configuration, all the expulsion forces are compensated 
($b)$ when $d=a$. The expulsion of the periodic system (c) is possible only at 
the relations $d/a>1$ and reaches its maximum at $d/a\sim 1.65$. The 
expulsion force grows linearly at increasing number of wings in the 
configuration.}}
\end{figure*}
Combined finding of maxima using all possible parameters according to 
formula (8) in Ref. \cite{Fateev:2012a} allows to find for the angle 
$\varphi \approx 1\mathring{ }$ optimal relations $d/a\to 1.65$ and $R/a\to 1.8$ 
for given geometry of the configuration of planes.

Let us find the number of nano-planes $n$, which can be placed along the length 
$Y=1\;\mbox{m}$ if the minimal distance between the wings is taken equal to 
the distance between the atoms of gold in the crystal lattice $a=4.0\times 
10^{-10}\;\mbox{m}$ \cite{Kittel:1995}. We find
\begin{equation}
\label{eq1}
n\approx \frac{Y}{d+a+2R\tan (\varphi /2)}.
\end{equation}
It means that for the optimal parameters
\[
n\approx 6.7\times 10^8.
\]
Assuming that the length of trapezoid cavity profiles is $L=1\;\mbox{m}$ and 
taking into account that each cavity has two wings, we find the propulsion 
force developed by a platform having an area of the order of 
$L\times Y=1\;\mbox{m}^2$
\begin{equation}
\label{eq2}
{\sum}F\approx 6\times 10^9 \, \mbox{newton}.
\end{equation}
Certainly, this value ${\sum}F$ is obtained for structures in certain ideal 
conditions and with rough approximations (however, in some respects the 
approximations are not rougher than those made by Casimir) which we have 
made developing the theory of the calculation of expulsion forces. It seems 
that we need several tens or even hundreds of layers with such structures to 
develop such propulsive force. It is still difficult to determine the exact 
number of the layers. However, we are going to make approximate estimates 
based on the obtained value ${\sum}F$.

We have received a pretty large value for the propulsive force. The mass of 
a body which can be balanced by such force (at the acceleration of free fall 
on the surface of the Earth $g\approx 9.8$\, m/s$^2$) is
\begin{equation}
\label{eq3}
M\approx \frac{{\sum}F}{g}=6.5\times 10^5 \,\mbox{tons}. 
\end{equation}
It means that Cheops pyramid ($6.25\times 10^6$ tons) will be balanced in 
the gravity force field by about 10\, m $^{2}$ of the monoatomic layer of our 
platforms!

Now let us estimate the amount of ideal metal (let it be gold) which is 
necessary for lifting and giving acceleration to all people inhabiting our 
planet. Let us assume that for one person a platform with the propulsive 
force capable of balancing 200 kg is sufficient. First, let us find what the 
area of single-layer (monoatomic layer) gold should be for all people. Let 
us calculate it using the mass of Cheops pyramid. If the mass of an average 
person is 70 kg and the total number of people is $7\times 10^9$, we obtain 
that all the people on the planet weigh about 78 masses of Cheops pyramids! 
It means that we need only about 780 m$^{2}$ of Casimir platforms for all 
the people for balancing in the Earth gravity field. And we need thrice as 
much (2400 m$^{2})$ if we need to fly in personal capsules wherever we want 
and with whatever speed. Let us also take into consideration that the 
platform consists of cavities; each cavity has two wings of the length 
$R\approx 1.8\times a$. Thus, the area of the platform surfaces is 
determined as follows:
\begin{equation}
\label{eq4}
S\approx 2n\,R\,2400\approx 4500\,\mbox{m}^2.
\end{equation}
How much will these meters of gold weigh? If $s=1\,m^2$ of a single-layer of 
gold with the lattice spacing $l\approx a=4.0\times 10^{-10}\;\mbox{m}$ and 
the mass of one gold atom $m_{Au} \approx 3.15\times 10^{-25}\mbox{kg}$ 
\cite{Kittel:1995} can weigh
\begin{equation}
\label{eq5}
m\approx \frac{s}{l^2}m_{Au} \approx 1.6\times 10^{-6}\,\mbox{kg}
\end{equation}
we need 
\[
M\approx S\;m\approx 0.01\;\mbox{kg}.
\]
This mass seems extremely small! However, if we calculate for 1000 
monoatomic layers, we will obtain $10$\,{kg} for all human beings! 
Seems very little! Well, it is so!

Why do we unexpectedly speak of 1000 layers in platforms? We will need that 
many, most likely, because of the existence of various unexpected physical 
factors preventing ``squeezing'' maximally possible expulsion forces. For 
example, the imperfection of the reflecting surface of the wings which will 
cause the decrease of directed recoil of photon momenta. Too thin wings can 
not allow to ``squeeze'' even 1{\%} of energy of virtual photons due to 
their transparency for too active virtual photons. Probably, there are some 
other problems as well \cite{Bordag:2009}. I hope that all possible problems can be 
solved in the nearest future.

If one layer of quantum structures is thinner than 10 angstroms, a 
many-layered ``pie'' of structures for the lift and flight of the object in 
the Universe, the mass of which is congruent to the mass of Cheops pyramid, 
will not exceed 10-100 angstroms. It is smaller than the diameter of a human 
hair by a factor of 1000! It is not much for the quantum flight of the 
pyramid, is it? And for the flight of one person, less than 10 square 
millimeters of one-layer structures will be necessary with regard to all 
possible corrections!

The author is grateful to T. Bakitskaya for his helpful participation in discussions.

\end{document}